\documentclass[amstex,12pt,thmsa,sw20bams]{article}%
\usepackage{amsfonts}
\usepackage[utf8]{inputenc}
\usepackage{graphicx}
\usepackage[francais]{babel}
\usepackage{amsmath}
\usepackage{amssymb}%
\providecommand{\U}[1]{\protect\rule{.1in}{.1in}}
\ifx\pdfoutput\relax\let\pdfoutput=\undefined\fi
\newcount\msipdfoutput
\ifx\pdfoutput\undefined\else
\ifcase\pdfoutput\else
\ifx\paperwidth\undefined\else
\ifdim\paperheight=0pt\relax\else\pdfpageheight\paperheight\fi
\ifdim\paperwidth=0pt\relax\else\pdfpagewidth\paperwidth\fi
\fi\fi\fi
\begin{document}

\author{Choulakian V, Universit\'{e} de Moncton, Canada
\and vartan.choulakian@umoncton.ca}
\title{Distortion in Correspondence Analysis and in Taxicab Correspondence Analysis:
A Comparison }
\date{March 2024}
\maketitle

\begin{abstract}
Distortion is a fundamental well-studied topic in dimension reduction papers,
and intimately related with the underlying intrinsic dimension of a mapping of
a high dimensional data set onto a lower dimension. In this paper, we study
embedding distortions produced by Correspondence Analysis and its robust
$l_{1}$ variant Taxicab Correspondence analysis, which are visualization
methods for contingency tables. For high dimensional data, distortions in
Correspondence Analysis are contractions; while distortions in Taxicab
Correspondence Analysis could be contractions or stretchings. This shows that
Euclidean geometry is quite rigid, because of the orthogonality property;
while Taxicab geometry is quite flexible, because the orthogonality property
is replaced by the conjugacy property.



\end{abstract}

\section{\textbf{Introduction}}

There are two\ kinds of dimensionality reduction clearly mentioned in Freksen
(2021): First, Johnson-Lindenstrauss type where data are projected using
random projections; second, PCA type where data are projected using optimal
projections. In our case correspondence analysis (CA) and taxicab CA (TCA) are
PCA type.

Distortion is a fundamental concept well-studied in dimension reduction
papers, and intimately related with the underlying intrinsic dimension of a
mapping of a high dimensional data set onto a lower dimension; see in
particular, Johnson and Lindenstrauss (1982, 1984), Bourgain (1985), Vankadara
and von Luxburg (2018), Freksen (2021), Agrawal et al. (2021) and Meila and
Zhang (2023).

Vankadara and von Luxburg (2018) define the commonly used notion of
\textit{global distortion} as \textquotedblright Distortion measures aim to
quantify the deviation of an embedding from isometry\textquotedblright.

Meila and Zhang (2023, subsection 5.1) define \textit{local distortion}, where
they state : \textquotedblright Many embedding algorithms tend to
\textit{contract} regions of M (manifold) where the data are densely sampled
and \textit{to stretch} the sparsely sampled regions.\textquotedblright\ This
observation seems to apply to TCA too but in visualization; see the Taxicab
parabola of type 1 in Figure 2 representing the structure of the data set in
Example 1. We note that there are 2 types of parabolas in Taxicab geometry,
see Figures 2 and 4.

The \textit{intrinsic}$\ $\textit{dimension} is an essential number for data
visualization and interpretation in PCA type analyses. We propose a new simple
criterion within TCA.

In this paper, we compare essential properties of distortion of the row or
column profiles of a contingency table from their barycenters. This will help
us also to deduce the lower and upper bounds of TCA intrinsic dimension of a
data set.

So the contents of this paper are: In section 2, we present an outline of CA
and TCA; section 3 presents main results; section 4 discusses applications;
and we conclude in section 5.

Benz\'{e}cri (1973, volumes 1 and 2) is the reference book on CA. Among
others, Greenacre (1984) is one of the best representative of Benz\'{e}cri's
approach; Beh and Lombardo (2014) present a panoramic review of CA. Taxicab CA
is a robust $l_{1}$ variant of CA, see Choulakian (2006, 2016, 2021).

\section{Preliminaries}

Let $\mathbf{N}=(n_{ij})$\textbf{ }for $i=1,...,I$ and $j=1,...,J$ be a 2-way
contingency table and $\mathbf{P=N/}n=(p_{ij})$ of size $I\times J$ the
associated correspondence matrix (probability table) of \textbf{N}, where
$n=\sum_{i,j}n_{ij}$. We define as usual $p_{i+}=\sum_{j=1}^{J}p_{ij}$ the
marginal probability of the $i$th row, and similarly $p_{+j}=\sum_{i=1}%
^{I}p_{ij}$ the marginal property of the $j$th column.

The CA association index of \textbf{P} is%

\begin{align*}
\Delta_{ij}  &  =(\frac{p_{ij}}{p_{i+}p_{+j}}-1)\\
&  =\frac{1}{p_{i+}p_{+j}}(p_{ij}-p_{i+}p_{+j})\\
&  =\frac{1}{p_{i+}}(\frac{p_{ij}}{p_{+j}}-p_{i+})\\
&  =\frac{1}{p_{+j}}(\frac{p_{ij}}{p_{i+}}-p_{+j}),
\end{align*}
where: $\frac{p_{ij}}{p_{i+}p_{+j}}$ is density function of $p_{ij}$ with
respect to $p_{i+}p_{+j},$ $\frac{p_{ij}}{p_{+j}}$ is the profile of the $j$th
column and similarly $\frac{p_{ij}}{p_{i+}}$ is the profile of the $i$th row.

The CA or Taxicab CA (TCA) decomposition is, for $k=$\bigskip$rank(p_{ij}%
-p_{i+}p_{+j}),$%

\[
\frac{p_{ij}}{p_{i+}p_{+j}}-1=\sum_{\alpha=1}^{k}f_{\alpha}(i)g_{\alpha
}(j)/\delta_{\alpha},
\]
calculated by generalized singular value decomposition (SVD) or taxicab SVD
(TSVD) of $\Delta_{ij}.$

A reference for the CA and TCA basic equations is Choulakian et al. (2014).

\subsection{CA basic equations}

The parameters $(f_{\alpha}(i),g_{\alpha}(j),\delta_{\alpha})$ satisfy: for
$\alpha,\beta=1,...,k=$\bigskip$rank(p_{ij}-p_{i+}p_{+j}):$

$\delta_{\alpha}^{2}=\sum_{i=1}^{I}f_{\alpha}^{\ \ 2}(i)p_{i+}=\sum_{j=1}%
^{J}g_{\alpha}^{2}(j)p_{+j}$

$0=\sum_{i=1}^{I}f_{\alpha}(i)p_{i+}=\sum_{j=1}^{J}g_{\alpha}(j)p_{+j}$

$0=\sum_{i=1}^{I}f_{\alpha}(i)f_{\beta}(i)p_{i+}=\sum_{j=1}^{J}g_{\alpha
}(j)g_{\beta}(j)p_{+j}$ \ \ for $\alpha\neq\beta.$

The last equatins show that the row factor scores $(f_{\alpha}(i))$ and the
column factor scores $(g_{\alpha}(j))$ for $\alpha=1,...,k$ \ are ORTHOGONAL;
which is an essential property of the EUCLIDEAN geometry.

The distance (also named Benz\'{e}cri distance) between the $i$th row profile
and its barycenter is

$dist_{Benz}^{2}(\frac{P_{ij}}{P_{i+}},P_{+j})=\sum_{j=1}^{J}\frac
{(\frac{P_{ij}}{P_{i+}}-P_{+j})^{2}}{P_{+j}}=\sum_{j=1}^{J}P_{+j}(\frac
{P_{ij}}{P_{+j}P_{i+}}-1)^{2}=\sum_{\alpha=1}^{k}(f_{\alpha}(i))^{2}.$

Similarly, the distance between the $j$th column profile and its barycenter is

$dist_{Benz}^{2}(\frac{P_{ij}}{P_{+j}},P_{i+})=\sum_{i=1}^{I}\frac
{(\frac{P_{ij}}{P_{+j}}-P_{i+})^{2}}{P_{i+}}=\sum_{i=1}^{I}P_{i+}(\frac
{P_{ij}}{P_{+j}P_{i+}}-1)^{2}=\sum_{\alpha=1}^{k}(g_{\alpha}(j))^{2}.$

The total variance (also named inertia) is:

$weightedAve=\sum_{i,j}\frac{(p_{ij}-p_{i+}p_{+j})^{2}}{p_{i+}p_{+j}}%
=\sum_{i,j}P_{i+}P_{+j}(\frac{P_{ij}}{P_{+j}P_{i+}}-1)^{2}=$

$\ \ =\sum_{i=1}^{I}P_{i+}\ dist_{Benz}^{2}(\frac{P_{ij}}{P_{i+}},P_{+j}%
)=\sum_{=1}^{J}P_{+j}\ dist_{Benz}^{2}(\frac{P_{ij}}{P_{+j}},P_{i+}%
)=\sum_{\alpha=1}^{k}\delta_{\alpha}^{2}.$

\subsection{TCA basic equations}

The parameters $(f_{\alpha}(i),g_{\alpha}(j),\delta_{\alpha})$ satisfy: for
$\alpha,\beta=1,...,k=$\bigskip$rank(p_{ij}-p_{i+}p_{+j}):$

$\delta_{\alpha}=\sum_{i=1}^{I}|f_{\alpha}^{\ }(i)|p_{i+}=\sum_{j=1}%
^{J}|g_{\alpha}(j)|p_{+j}$

$0=\sum_{i=1}^{I}f_{\alpha}(i)p_{i+}=\sum_{j=1}^{J}g_{\alpha}(j)p_{+j}$

$0=\sum_{i=1}^{I}f_{\alpha}(i)\ sign(f_{\beta}(i))p_{i+}=\sum_{j=1}%
^{J}g_{\alpha}(j)\ sign(g_{\beta}(j))p_{+j}$\ \ for $\alpha>\beta.$

The last equatins show that the row factor scores $(f_{\alpha}(i))$ and the
column factor scores $(g_{\alpha}(j))$ for $\alpha=1,...,k$ \ are CONJUGATE;
which is an essential property of the TAXICAB geometry.

The taxicab distance between the $i$th row profile and its barycenter is

$dist_{Taxi}(\frac{P_{ij}}{P_{i+}},P_{+j})=\sum_{j=1}^{J}P_{+j}|\frac{P_{ij}%
}{P_{+j}P_{i+}}-1|=\sum_{j=1}^{J}|(\frac{P_{ij}}{P_{i+}}-P_{+j})|\ \leq
\sum_{\alpha=1}^{k}|f_{\alpha}(i)|.$

Similarly, the taxicab distance between the $j$th column profile and its
barycenter is

$distd_{Taxi}(\frac{P_{ij}}{P_{j+}},P_{i+})=\sum_{i=1}^{I}P_{i+}|\frac{P_{ij}%
}{P_{+j}P_{i+}}-1|=\sum_{i=1}^{I}|(\frac{P_{ij}}{P_{j+}}-P_{i+})|\ \leq
\sum_{\alpha=1}^{k}|g_{\alpha}(j)|.$

The total raw data dispersion is

$weightedAve=\sum_{i,j}|P_{ij}-P_{i+}P_{+j}|\ =\sum_{i,j}P_{i+}P_{+j}%
|\frac{P_{ij}}{P_{+j}P_{i+}}-1|\ =$

$\ \ \ \ =\sum_{j=1}^{J}P_{+j}\ dist_{Taxi}(\frac{P_{ij}}{P_{+j}},P_{i+}%
)=\sum_{i=1}^{I}P_{i+}\ dist_{Taxi}(\frac{P_{ij}}{P_{i+}},P_{+j})\leq
\sum_{\alpha=1}^{k}\delta_{\alpha}.$

\section{CA and TCA distortions}

\textbf{Distortion}: Comparison of the distances on the raw data set and on
the maps. Given two metric spaces $(X,dist_{x})$ and $(Y,dist_{y})$, and an
injective mapping $\phi:X->Y,$ for two points $(u,v)\in X\times X$ and their
mappings $(\phi(u),\phi(v))\in Y\times Y$, our aim is the comparison of the
distances $dist_{x}(u,v)$ and $dist_{y}(\phi(u),\phi(v))$. We can have 3 cases:

\textbf{Case 1}: $dist_{x}(u,v)=dist_{y}(\phi(u),\phi(v))$; distortion is
null, and the mapping $\phi$ is called \textit{isometric embedding} for the
pair of points $(u,v)$.

\textbf{Case 2}: $dist_{x}(u,v)>dist_{y}(\phi(u),\phi(v))$; distortion is a
contraction, and the mapping $\phi$ is called \textit{contractive embedding}
for the pair of points $(u,v)$.

\textbf{Case 3}: $dist_{x}(u,v)<dist_{y}(\phi(u),\phi(v))$; distortion is a
stretching (also named an expansion), and the mapping $\phi$ is called
\textit{stretch embedding} for the pair of points $(u,v)$.

Let $d$ be the embedding dimension. For the next two subsections, we accept
the following:

\textbf{Assumption: }For\textbf{ }$d=1,...,rank(p_{ij}-p_{i+}p_{+j})$,
\ \textbf{ }$\sum_{\alpha=1}^{d}|$\textbf{ }$f_{\alpha}(i)|\ \neq0.$

\subsection{CA distortion}

\textbf{In\ CA}: distortion is a CONTRACTION.

For $1\leq d<rank(p_{ij}-p_{i+}p_{+j})$

a) $dist_{Benz}^{2}(\frac{P_{ij}}{P_{i+}},P_{+j})=\sum_{j=1}^{J}\frac
{(\frac{P_{ij}}{P_{i+}}-P_{+j})^{2}}{P_{+j}}=\sum_{j=1}^{J}P_{+j}(\frac
{P_{ij}}{P_{+j}P_{i+}}-1)^{2}\geq\sum_{\alpha=1}^{d}(f_{\alpha}(i))^{2}$

b) $weightedAve=\sum_{i=1}^{I}P_{i+}\ dist_{Benz}^{2}(\frac{P_{ij}}{P_{i+}%
},P_{+j})=\sum_{i,j}\frac{(p_{ij}-p_{i+}p_{+j})^{2}}{p_{i+}p_{+j}}%
>\sum_{\alpha=1}^{d}\delta_{\alpha}^{2}$

This can be re-expressed in the following way: There exists a constant
$c_{1}\in(0,1)$ such that

$c_{1}$ $dist_{Benz}^{2}(\frac{P_{ij}}{P_{i+}},P_{+j})\leq\sum_{\alpha=1}%
^{d}(f_{\alpha}(i))^{2}\leq dist_{Benz}^{2}(\frac{P_{ij}}{P_{i+}},P_{+j})$

or

$c_{1}\leq\frac{\sum_{\alpha=1}^{d}(f_{\alpha}(i))^{2}}{dist_{Benz}^{2}%
(\frac{P_{ij}}{P_{i+}},P_{+j})}\leq1$

where $c_{1}=\min_{i}\frac{\sum_{\alpha=1}^{d}(f_{\alpha}(i))^{2}}%
{dist_{Benz}^{2}(\frac{P_{ij}}{P_{i+}},P_{+j})}$

\subsection{TCA distortion}

\textbf{In TCA}: distortion is CONTRACTION or STRETCHING (also named
EXPANSION). The sign $?$ that we use below, means equal, less than or greater than.

a1) For $d=1$

$dist_{Taxi}(\frac{P_{ij}}{P_{i+}},P_{+j})=\sum_{j=1}^{J}|(\frac{P_{ij}%
}{P_{i+}}-P_{+j})|\ \geq|f_{1}(i)|\ $is a CONTRACTION

a2) For $2\leq d<rank(p_{ij}-p_{i+}p_{+j})$

$dist_{Taxi}(\frac{P_{ij}}{P_{+j}},P_{i+})=\sum_{j=1}^{J}|(\frac{P_{ij}%
}{P_{i+}}-P_{+j})|\ =\sum_{j=1}^{J}P_{+j}|\frac{P_{ij}}{P_{+j}P_{i+}%
}-1|\ ?\sum_{\alpha=2}^{d}|f_{\alpha}(i)|\ \ $\ $\ \ $

b1) For $d=1$

$weightedAve=\sum_{j=1}^{J}P_{+j}\ dist_{Taxi}(\frac{P_{ij}}{P_{+j}}%
,P_{i+})=\sum_{i,j}|P_{ij}-P_{i+}P_{+j}|\ >\delta_{1}$

b2) For $2\leq d<rank(p_{ij}-p_{i+}p_{+j})$

$weightedAve=\sum_{j=1}^{J}P_{+j}\ dist_{Taxi}(\frac{P_{ij}}{P_{+j}}%
,P_{i+})=\sum_{i,j}|P_{ij}-P_{i+}P_{+j}|\ ?\ \sum_{\alpha=1}^{d}{}%
\delta_{\alpha}$

This can be re-expressed in the following way: There exist 2 constants
$c_{1}>0,$ $c_{2}>1$ and $c_{2}>c_{1}$, such that

$c_{1}$ $dist_{Taxi}(\frac{P_{ij}}{P_{+j}},P_{i+})\leq\sum_{\alpha=1}%
^{d}|f_{\alpha}(i)|\leq c_{2}$ $dist_{Taxi}(\frac{P_{ij}}{P_{+j}},P_{i+}%
)$\bigskip

or

$c_{1}\leq\frac{\sum_{\alpha=1}^{d}|f_{\alpha}(i)|}{dist_{Taxi}(\frac{P_{ij}%
}{P_{+j}},P_{i+})}\leq c_{2}$

where

$c_{1}=\min_{i}\frac{\sum_{\alpha=1}^{d}|f_{\alpha}(i)|}{dist_{Taxi}%
(\frac{P_{ij}}{P_{+j}},P_{i+})}$ and $c_{2}=\max_{i}\frac{\sum_{\alpha=1}%
^{d}|f_{\alpha}(i)|}{dist_{Taxi}(\frac{P_{ij}}{P_{+j}},P_{i+})}$

\textbf{Remark}: We see that for the embedding dimension $d=1$, the distortion
is a CONTRACTION for both CA and TCA. While, for $d\geq2$, the nature of
distortion can vary between CA and TCA.

We have:

\textbf{Definition}:

a) The upper bound for $TCA\ intrinsic\ dimension$ is:

\ \ \ \ \ \ \ $\dim Upper_{intrinsic}=\arg\min_{d}(\sum_{\alpha=1}^{d}%
\delta_{\alpha}-|\sum_{i,j}|P_{ij}-P_{i+}P_{+j}|).$

b) The lower bound for $TCA\ intrinsic\ dimension$ is:

\ \ \ \ \ \ \ $\dim Lower_{intrinsic}=\arg\min_{d}(\sum_{\alpha=1}^{d}%
\sum_{i,j}|P_{ij}-P_{i+}P_{+j}|\ -\delta_{\alpha}).$

\textbf{Corollary}: By b1) $\dim Upper_{intrinsic}\geq2$ and $\dim
Lower_{intrinsic}\geq1.$

\section{Examples}

We present some comparisons of the main results presented in the precedent
section. We apply three R packages: the \textit{ca} package by Greenacre et
al. (2022 ), the \textit{ade4} package by Dray et al. (2023) and
\textit{TaxicabCA} by Allard and Choulakian (2019).

\subsection{Colors of music data set}

The data set is a contingency table of size $10\times9$, found in Abdi and
Bera (2016); where $J=9$ kinds of music and $I=10$ colors. It is constructed
in the following way: \textquotedblright A set of 12 children and 10 adults
picked up the color that best describes each of 9 pieces of
music\textquotedblright. The sparsity of the data set is 21.11\% using the
following R code:

sum(dataMatrix == 0)/length(dataMatrix)

[1] 0.2111111

Figures 1 and 2 display the CA and TCA maps; we note some important
differences. In the CA plot, the Black color dominates first principal
dimension. In Figure 2, the rows and the columns represent taxicab parabolas,
where the absolute value of the slope of the directrix is greater than 1; see
among others Gardner (1997, ch. 10), Wilson (2011) and Petrovi\'{c} et al.
(2014). The first principal dimension has very simple interpretation: From the
left side to right, the colors are ordered rom the \textit{most concentrated}
(\textit{Black} and \textit{Brown}) to the \textit{most dispersed}
(\textit{purple} and \textit{pink}); see our comments on Meila and Zhang
(2023, subsection 5.1) in the introduction. The same interpretation applies
also to the types of music.

\begin{figure}[h]
\label{fig:MagEngT}
\centering 
\includegraphics[scale=1.3]{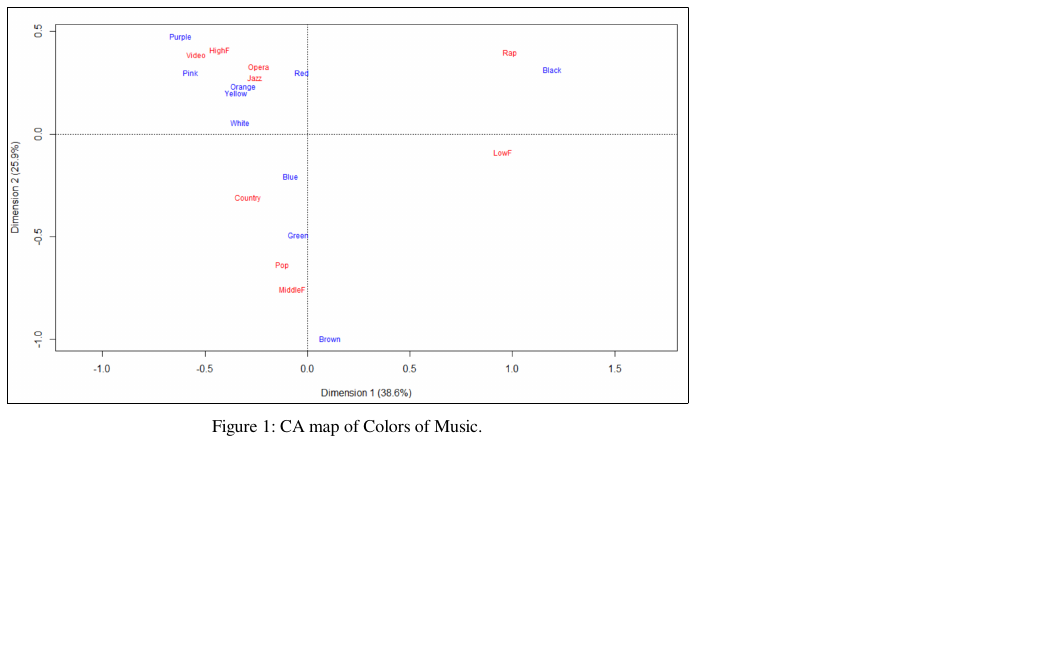}  
\end{figure}

\begin{figure}[h]
\label{fig:MagEngT}
\centering 
\includegraphics[scale=1.3]{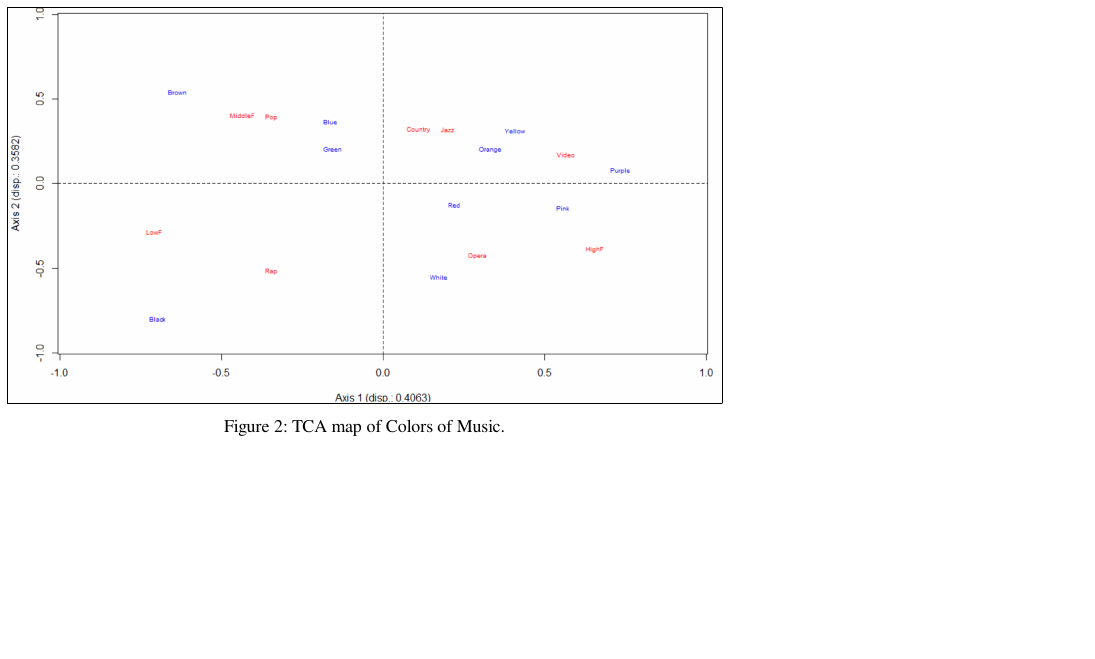}  
\end{figure}

Table 1 displays some important CA raw data and embedded distance values of
the 10 colors: $\sum_{j=1}^{J}\frac{(\frac{P_{ij}}{P_{i+}}-P_{+j})^{2}}%
{P_{+j}}=$ $dist_{Benz}^{2}(\frac{P_{ij}}{P_{i+}},P_{+j})$ is the Benz\'{e}cri
distance between the $i$th color profile and its barycenter; while for the
embedding dimensions $d=1,2,3,$ $\sum_{\alpha=1}^{d}|f_{\alpha}(i)|^{2}%
=dist_{d}^{2}(\phi(\frac{P_{ij}}{P_{i+}}),\phi(P_{+j}))$ where $\phi=CA$
embedding on the first $d$ principal dimensions. We see that $dist_{Benz}%
^{2}(\frac{P_{ij}}{P_{i+}},P_{+j})>dist_{d}^{2}(\phi(\frac{P_{ij}}{P_{i+}%
}),\phi(P_{+j})),$ so distortion values resulting from the CA embedddings on
the first three principal dimensions for the 10 rows (colors) represent contractions.

\bigskip%
\begin{tabular}
[c]{l|l|l|l|l|}%
\multicolumn{5}{l|}{\textbf{Table 1: CA distances of the 10 colors from their
barycenter.}}\\\hline\hline
\multicolumn{1}{|l|}{\textit{color}$(i)$} & $dist_{Benz}^{2}(\frac{P_{ij}%
}{P_{i+}},P_{+j})$ & $|f_{1}(i)|^{2}$ & $\sum_{\alpha=1}^{2}|f_{\alpha
}(i)|^{2}$ & $\sum_{\alpha=1}^{3}|f_{\alpha}(i)|^{2}$\\\hline
Red & 0.2187 & $0.0007$ & $0.0904$ & $0.1054$\\
Orange & 0.3333 & $0.0984$ & $0.11521$ & $0.2099$\\
Yellow & 0.4544 & $0.1211$ & $0.1618$ & $0.2597$\\
Green & 0.4121 & $0.0019$ & $0.2423$ & $0.0490$\\
Blue & 0.5208 & $0.0067$ & $0.0490$ & $0.1001$\\
Purple & 0.7574 & $0.3828$ & $0.6082$ & $0.6104$\\
White & 1.3878 & $0.1074$ & $0.1106$ & $1.2299$\\
Black & 1.5363 & $1.4275$ & $1.5270$ & $1.5316$\\
Pink & 0.8750 & $0.3249$ & $0.4152$ & $0.5629$\\
Brown & 1.0204 & $0.0127$ & $1.0059$ & $1.0139$\\\hline\hline
weightedAve & 0.7462 & $0.2880$ & $0.48113$ & $0.6196$\\
cols\&rows & $\sum_{i,j}\frac{(p_{ij}-p_{i+}p_{+j})^{2}}{p_{i+}p_{+j}}$ &
$\delta_{1}^{2}$ & $\delta_{1}^{2}+\delta_{2}^{2}$ & $\sum_{\alpha=1}%
^{3}\delta_{\alpha}^{2}$\\\hline\hline
\end{tabular}

Table 2 displays some important TCA raw data and embedded distance values of
the 10 colors: $\sum_{j=1}^{J}|\frac{P_{ij}}{P_{i+}}-P_{+j}|=$ $dist_{Taxi}%
(\frac{P_{ij}}{P_{i+}},P_{+j})$ is the Taxicab distance between the $i$th
color profile and its barycenter; while for $d=1,2,3,$ $\sum_{\alpha=1}%
^{d}|f_{\alpha}(i)|\ =dist_{d}(\phi(\frac{P_{ij}}{P_{i+}}),\phi(P_{+j}))$
where $\phi=TCA$ embedding on the first $d$ principal dimensions. We see that
for $d=1,\ dist_{Taxi}(\frac{P_{ij}}{P_{i+}},P_{+j})>dist_{1}(\phi
(\frac{P_{ij}}{P_{i+}}),\phi(P_{+j})),$ so distortion values resulting from
the TCA embedddings on the first principal dimension for the 10 rows (colors)
represent contractions; while for $d=2$, $dist_{Taxi}(\frac{P_{ij}}{P_{i+}%
},P_{+j})?dist_{2}(\phi(\frac{P_{ij}}{P_{i+}}),\phi(P_{+j})),$ so distortion
values resulting from the TCA embedddings on the first two principal
dimensions for the 10 rows (colors) represent contractions or stretchings (in
bold colors). Finally, for $d=3$, $dist_{Taxi}(\frac{P_{ij}}{P_{i+}}%
,P_{+j})<dist_{3}(\phi(\frac{P_{ij}}{P_{i+}}),\phi(P_{+j})),$ so distortion
values resulting from the TCA embedddings on the first three principal
dimensions for the 10 rows (colors) represent stretchings (in bold colors).

Furthermore, $TCA\ intrinsic\ dimension$ is $\dim_{intrinsic}=2.$%

\begin{tabular}
[c]{l|l|l|l|l|}%
\multicolumn{5}{l|}{\textbf{Table 2: TCA distances of the 10 colors from their
barycenter.}}\\\hline\hline
\multicolumn{1}{|l|}{\textit{music}$(i)$} & $dist_{Taxi}(\frac{P_{ij}}{P_{+j}%
},P_{i+})$ & $|f_{1}(i)|$ & $\sum_{\alpha=1}^{2}|f_{\alpha}(i)|$ &
$\sum_{\alpha=1}^{3}|f_{\alpha}(i)|$\\\hline
Red  & 0.4444 & $0.2222$ & $0.3456$ & \textbf{0.4671}\\
Orange  & 0.4444 & $0.3333$ & \textbf{0.5371} & \textbf{0.7394}\\
Yellow  & 0.5688 & $0.4088$ & \textbf{0.7152} & \textbf{0.9890}\\
Green  & 0.5411 & $0.1546$ & $0.3564$ & \textbf{0.7083}\\
Blue  & 0.5965 & $0.1637$ & $0.5354$ & \textbf{0.7459}\\
Purple  & 0.8034 & $0.7350$ & \textbf{0.8141} & \textbf{0.9773}\\
White  & 1.0476 & $0.1746$ & $0.7239$ & \textbf{1.5496}\\
Black  & 1.0038 & $0.6973$ & \textbf{1.4940} & \textbf{1.9288}\\
Pink  & 0.7777 & $0.5555$ & $0.6974$ & \textbf{1.0278}\\
Brown  & 0.9206 & $0.6349$ & \textbf{1.1779} & \textbf{1.5834}\\\hline\hline
weightedAve & 0.7048 & $0.4063$ & \textbf{0.7644} & \textbf{1.0890}\\
cols\&rows & $\sum_{i,j}|P_{ij}-P_{i+}P_{+j}|$ & $\delta_{1}$ & $\delta
_{1}+\delta_{2}$ & $\sum_{\alpha=1}^{3}\delta_{\alpha}$\\\hline\hline
\end{tabular}

\subsection{Aravo ecological abundance data set}

We consider the \textit{Aravo} data set \textbf{N }of size $75\times82$ found
in the R package \textit{ade4 }by Dray et al.(2023). This dataset describes
the distribution of 82 species of Alpine plants in 75 sites. The percentage of
zero counts in \textbf{N} is $59.29\%$\textbf{.}

Figures 3 and 4 display the CA and TCA maps; both represent parabolas. However,
in Figure 4, the rows represent taxicab parabola, where the absolute value of
the slope of the directrix is less than 1. Figures 2 and 4\ represent the two
kinds of Taxicab parabolas as discussed by, see among others Gardner (1997, ch.
10), Wilson (2011) and Petrovi\'{c} et al. (2014).

we have: $\delta_{1}=0.627$\textbf{, }$\delta_{1}+\delta_{2}=1.085,\ \delta
_{1}+\delta_{2}+\delta_{3}=1.463$ and $\sum_{i,j}|P_{ij}-P_{i+}P_{+j}|=1.249;$
so $\dim Upper_{intrinsic}=3$ and $\dim Lower_{intrinsic}=2;$ given the
parabola, we deduce that TCA $\dim_{intrinsic}=2.$%

\begin{figure}[h]
\label{fig:MagEngT}
\centering 
\includegraphics[scale=1.3]{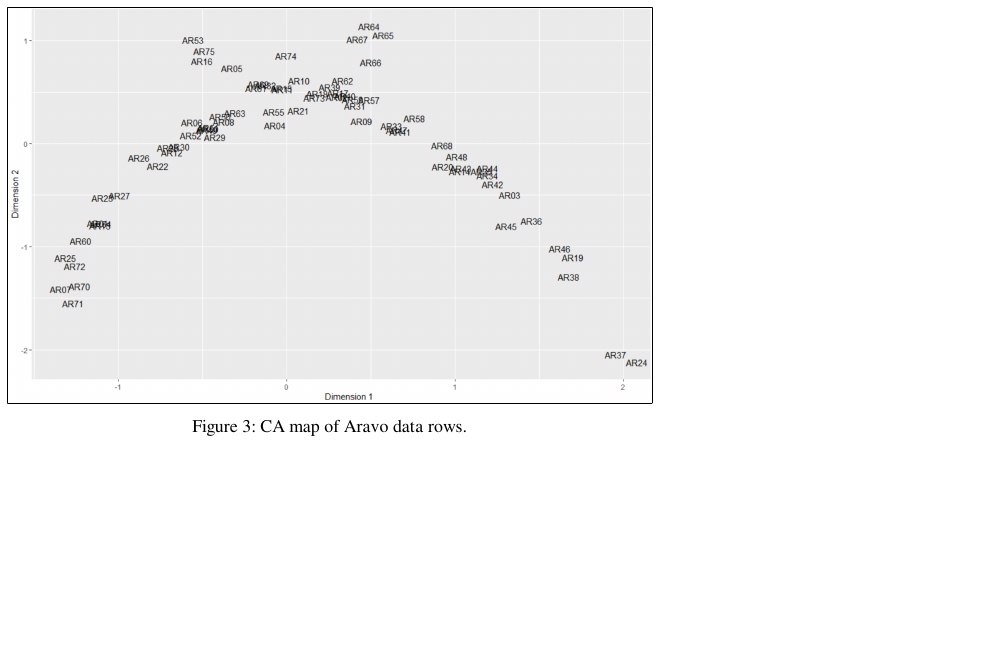}  
\end{figure}

\begin{figure}[h]
\label{fig:MagEngT}
\centering 
\includegraphics[scale=1.3]{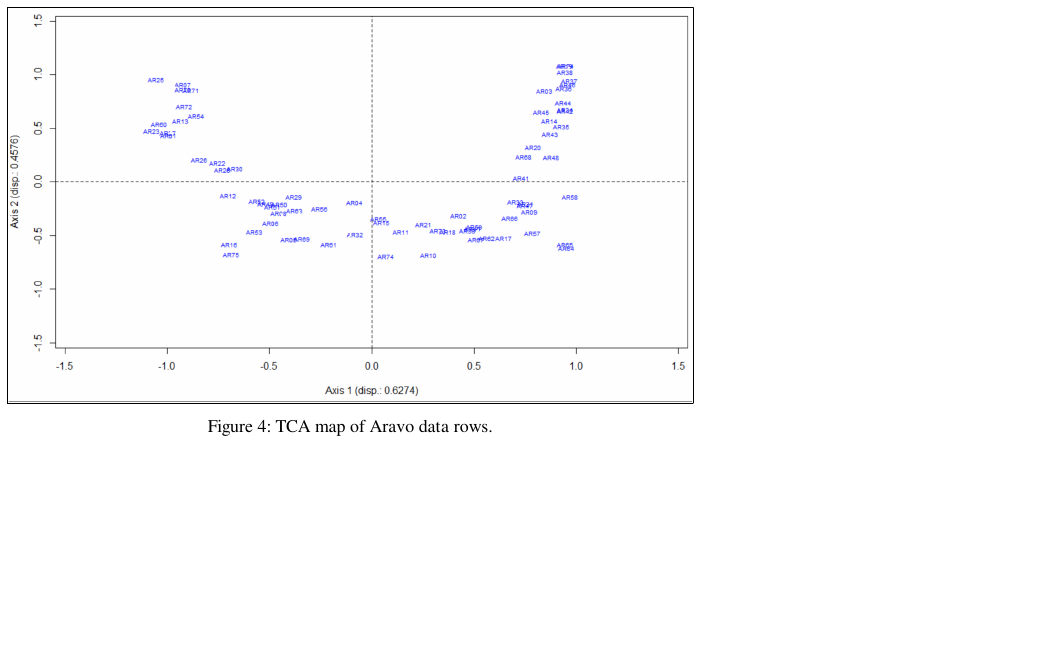}  
\end{figure}

\subsection{\textit{Rodent} ecological abundance data}

We consider the \textit{rodent} data set \textbf{N }of size $28\times9$ found
in the R package \textit{TaxicabCA }by Allard and Choulakian (2019), taken
from Quinn and Keough (2002). This is an abundance data set of 9 species of
rodents in 28 cities in California. The percentage of zero counts in
\textbf{N} is $66.27\%$\textbf{.} Choulakian (2017) analyzed it by comparing
the CA and TCA maps; using Benz\'{e}cri's criterion, Choulakian (2021) showed
that it has quasi 2-blocks diagonal structure (or partly decomposable);
furthermore, Choulakian (2023) analyzed it by Goodman's marginal-free
correspondence analysis.

We leave the interpretation of the distortion values in Tables 3 and 4 to the reader.

However we see that: $TCA\ intrinsic\ dimension$ is $\dim_{intrinsic}=2.$%

\begin{tabular}
[c]{l|l|l|l|l|}%
\multicolumn{5}{l|}{\textbf{Table 3: CA distances of the 9 rodents from their
barycenter.}}\\\hline\hline
\multicolumn{1}{|l|}{\textit{rodent}$(j)$} & $\sum_{i=1}^{I}\frac
{(\frac{P_{ij}}{P_{+j}}-P_{i+})^{2}}{P_{i+}}$ & $|g_{1}(j)|^{2}$ &
$\sum_{\alpha=1}^{2}|g_{\alpha}(j)|^{2}$ & $\sum_{\alpha=1}^{3}|g_{\alpha
}(j)|^{2}$\\\hline
\textit{1} & 34.91 & $6.79$ & $34.86$ & $34.87$\\
\textit{2} & 5.85 & $5.24$ & $5.84$ & $5.85$\\
\textit{3} & 0.245 & $0.09$ & $0.10$ & $0.12$\\
4 & 1.60 & $0.17$ & $0.17$ & $1.49$\\
5 & 0.49 & $0.10$ & $0.10$ & $0.11$\\
6 & 0.60 & $0.07$ & $0.08$ & $0.36$\\
\textit{7} & 3.00 & $0.20$ & $0.20$ & $2.04$\\
\textit{8} & 2.50 & $0.08$ & $0.09$ & $0.10$\\
9 & 1.63 & $0.15$ & $0.15$ & $0.35$\\\hline\hline
weightedAve & 1.719 & $0.746$ & $1.205$ & $1.493$\\
cols\&rows & $\sum_{i,j}\frac{(p_{ij}-p_{i+}p_{+j})^{2}}{p_{i+}p_{+j}}$ &
$\delta_{1}^{2}$ & $\delta_{1}^{2}+\delta_{2}^{2}$ & $\sum_{\alpha=1}%
^{3}\delta_{\alpha}^{2}$\\\hline\hline
\end{tabular}

\bigskip

\begin{tabular}
[c]{l|l|l|l|l|}%
\multicolumn{5}{l|}{\textbf{Table 4: TCA distances of the 9 rodents from their
barycenter.}}\\\hline\hline
\multicolumn{1}{|l|}{\textit{rodent}$(j)$} & $\sum_{i=1}^{I}|(\frac{P_{ij}%
}{P_{+j}}-P_{i+})|$ & $|g_{1}(j)|$ & $\sum_{\alpha=1}^{2}|g_{\alpha}(j)|$ &
$\sum_{\alpha=1}^{3}|g_{\alpha}(j)|$\\\hline
\textit{1} & 1.888 & $0.803$ & $1.595$ & \textbf{2.522}\\
\textit{2} & 1.408 & $0.876$ & \textbf{1.819} & \textbf{2.483}\\
\textit{3} & 0.390 & $0.306$ & \textbf{0.489} & \textbf{0.682}\\
4 & 1.15 & $0.848$ & \textbf{1.605} & \textbf{2.134}\\
5 & 0.542 & $0.149$ & $0.341$ & $0.514$\\
6 & 0.547 & $0.426$ & \textbf{0.814} & \textbf{1.124}\\
\textit{7} & 1.439 & $1.232$ & \textbf{2.110} & \textbf{2.689}\\
\textit{8} & 1.141 & $0.558$ & \textbf{1.621} & \textbf{2.425}\\
9 & 1.212 & $0.482$ & $0.527$ & \textbf{1.324}\\\hline\hline
weightedAve & 0.705 & $0.478$ & \textbf{0.900} & \textbf{1.248}\\
cols\&rows & $\sum_{i,j}|P_{ij}-P_{i+}P_{+j}|$ & $\delta_{1}$ & $\delta
_{1}+\delta_{2}$ & $\sum_{\alpha=1}^{3}\delta_{\alpha}$\\\hline\hline
\end{tabular}

\subsection{Sacred books data set}

The data set represents an extremely sparse contingency table of textual data
(a bag of words) found in Sah and Fokoue (2019). It is of size $I\times
J=590\times8265$, where $I=590$ represents fragments of chapters of 8 sacred
books and $J=8265$ represents the number of distinct words. The real
percentage of zero cells is 98.65\%. It is studied and visualized quite in
detail by (12+1) dimension reduction methods in Ma, Sun and Zou (2022,2023).
Choulakian and Allard (2023) visualized this data set by the joint use of CA
and TCA methods. From Table 1 in Choulakian and Allard (2023), we have the
first 4 principal values: $\delta_{1}=0.669$\textbf{, }$\delta_{2}%
=0.437,\delta_{3}=0.431,\delta_{4}=0.391$ and $\sum_{i,j}|P_{ij}-P_{i+}%
P_{+j}|=1.748976;$ so $\dim Upper_{intrinsic}=4$ and $\dim Lower_{intrinsic}%
=3.$

\subsection{Saporta-Tambrea data set}

Saporta and Tambrea (1993, II.3 example) present a data table of size
$19\times13$, where the counts represent the number of times each of 1000
respondents associates an item ( among 19) to 13 brands of diet butters. Due
to multiple answers, the total number of counts $n=21900$. There are no zero
counts in the table. By applying the Malinvaud test, they found the
$CA\ intrinsic\ dimension=2.$

Using TCA, we have: $\delta_{1}=0.05386276$\textbf{, }\ $\delta_{1}+\delta
_{2}=0.08952342,\ \delta_{1}+\delta_{2}+\delta_{3}=0.12062140$ and $\sum
_{i,j}|P_{ij}-P_{i+}P_{+j}|=0.08858070;$ so $TCA\ intrinsic\ dimension$ is
$\dim_{intrinsic}=2.$

\subsection{Food Of The World data set}

This data set is a contingency table of size $26\times68,$ where 26 world
cuisines are described by their $68$ cooking ingredients; it is available
online at

https://github.com/HerveAbdi/data4PCCAR/tree/master/data

The percentage of zero cells is 21.15\%.

CA and TCA maps (not shown) are quite similar. Using TCA, we have: $\delta
_{1}=0.4083893$\textbf{, }\ $\delta_{1}+\delta_{2}=0.6428507,\delta_{1}%
+\delta_{2}+\delta_{3}=0.8548540$ and $\sum_{i,j}|P_{ij}-P_{i+}P_{+j}%
|\ =0.5907166;$ so $TCA\ intrinsic\ dimension$ is $\dim_{intrinsic}=2.$

\section{Conclusion}

In dimension reduction methods for visualization problems, our impression from
observations is that EUCLIDEAN\ GEOMETRY is quite RIGID, because of the
ORTHOGONALITY property; so embedding distortions are contractions. While
TAXICAB\ GEOMETRY\ is quite FLEXIBLE, because orthogonality property is
replaced by the CONJUGACY\ property, thus embedding distortions are
contractions or stretchings. To be more precise as we stated in section 2:

In CA, the variance ordered factor scores are orthogonal:

$0=\sum_{i=1}^{I}f_{\alpha}(i)f_{\beta}(i)p_{i+}=\sum_{j=1}^{J}g_{\alpha
}(j)g_{\beta}(j)p_{+j}$ \ \ for $\alpha\neq\beta.$

In TCA, the dispersion ordered factor scores are conjugate:

$0=\sum_{i=1}^{I}f_{\alpha}(i)\ sign(f_{\beta}(i))p_{i+}=\sum_{j=1}%
^{J}g_{\alpha}(j)\ sign(g_{\beta}(j))p_{+j}$\ \ for $\alpha>\beta.$

To see clearly the flexibility of the Taxicab approach we provide two simple examples.

\textbf{Example 1}: We consider two vectors \textbf{v}$_{1}=(_{2}^{1})$ and
\textbf{v}$_{2}=(_{1}^{2}).$ It is IMPOSSIBLE to find a vector \textbf{v\ }%
$\mathbf{=\ }(_{y}^{x})$\textbf{\ }$\neq\mathbf{0}$ which is orthogonal to
both vectors \textbf{v}$_{1}$ and \textbf{v}$_{2}$; that is, \textbf{v}$^{t}%
$\textbf{v}$_{1}\ =\ $\textbf{v}$^{t}$\textbf{v}$_{2}=0$. While it is POSSIBLE
to find a vector \textbf{v} that is conjugate to both vectors \textbf{v}$_{1}$
and \textbf{v}$_{2}$: for instance \textbf{v\ }$=$ $(_{-3}^{3});$ then
\textbf{v}$^{t}$sign(\textbf{v}$_{1})=\ $\textbf{v}$^{t}$sign(\textbf{v}%
$_{2})=0$.

\textbf{Example 2}: The famous recreational mathematician Gardner (1997, ch.
10) provides many examples of Taxicab plane gometry, which distinguishes it
from the Euclidean plane geometry, and shows its flexibility. We provided an
example of data analysis displayed in Figure 2. A second simple example
discussed by Gardner is: \textquotedblright A taxicab \textit{scalene
triangle} with corners A, B, and C and sides of 14, 8, and 6 is shown at the
left in Figure 65. The sides of taxi polygons must of course be taxi paths,
and the paths that make up a polygon of specified dimensions may vary in shape
but not in length. Observe how the triangle in the illustration violates the
Euclidean theorem that the sum of any two sides of a triangle must be greater
than the third side. In this case the sum of two sides equals the third: 6 + 8
equals 14.\textquotedblright

As stated in Example 2, Taxicab plane geometry is considered recreational
mathematics and a curiosity for high-school students. But we think, it is a
new open rich area of research for data analysis worth pursuing. $\bigskip$

\textit{Acknowledgements: }Choulakian's research has been supported by NSERC
of Canada.\bigskip

\textbf{References}

Abdi H, B\'{e}ra M (2017) Correspondence Analysis. In: Alhajj, R., Rokne, J.
(eds) \textit{Encyclopedia of Social Network Analysis and Mining}. Springer,
New York, NY. https://doi.org/10.1007/978-1-4614-7163-9\_140-2

Agrawal A, Ali A, Boyd S (2021) \textit{Minimum-Distortion Embedding}.
Available at: https://arxiv.org/pdf/2103.02559.pdf

Allard J, Choulakian V (2019) \textit{Package TaxicabCA in R}

Beh E, Lombardo R (2014) \textit{Correspondence Analysis: Theory, Practice and
New Strategies}. N.Y: Wiley

Benz\'{e}cri JP (1973)\ \textit{L'Analyse des Donn\'{e}es: Vol. 1: La
Taxinomie}. Paris: Dunod

Benz\'{e}cri JP (1973)\ \textit{L'Analyse des Donn\'{e}es: Vol. 2: L'Analyse
des Correspondances}. Paris: Dunod

Bourgain J (1985). On Lipschitz embedding of finite metric spaces in Hilbert
space. \textit{Israel Journal of Mathematics}, 52(1-2):46--52

Choulakian V (2006) Taxicab correspondence analysis. \textit{Psychometrika,}
71, 333-345

Choulakian V (2016) Matrix factorizations based on induced norms.
\textit{Statistics, Optimization and Information Computing}, 4, 1-14

Choulakian V (2017) Taxicab correspondence analysis of sparse contingency
tables. \textit{Italian Journal of Applied Statistics,} 29 (2-3), 153-179

Choulakian V (2021) Quantification of intrinsic quality of a principal
dimension in correspondence analysis and taxicab correspondence analysis.
Available on \textit{arXiv:2108.10685}

Choulakian V (2023) Scale-free correspondence analysis. Available at:

\ \ \ \textit{https://arxiv.org/pdf/2311.17594.pdf}

Choulakian V, Simonetti B, Pham Gia, T (2014) Some new aspects of taxicab
correspondence analysis. \textit{Statistical Methods and Applications}, 23, 401--416

Choulakian V, Allard J (2023) Visualization of extremely sparse contingency
table by taxicab correspondence analysis : A case study of textual data.
Available at \ \textit{https://arxiv.org/pdf/2308.03079.pdf}

St\'{e}phane D, Dufour AB, Thioulouse J (2023) \textit{Package ade4 in R}

Freksen CB (2021) An Introduction to Johnson--Lindenstrauss Transforms.
Available at: \textit{https://arxiv.org/pdf/2103.00564.pdf}

Gardner M (1997) \textit{The Last Recreations}. Springer-Verlag New York, Inc.

Greenacre M (1984) \textit{Theory and Applications of Correspondence
Analysis}. Academic Press

Greenacre M, Nenadic O, Friendly M (2022) \textit{Package ca in R}

Johnson WB, Lindenstrauss J (1984), Extensions of Lipschitz mappings into a
Hilbert space. In Beals, Richard; Beck, Anatole; Bellow, Alexandra; et al.
(eds.), \textit{Conference in modern analysis and probability} (New Haven,
Conn., 1982), \textit{Contemporary Mathematics}, vol. 26, Providence, RI:
American Mathematical Society, pp. 189--206

Petrovi\'{c} M, Male\v{s}evi\'{c} B, Banjac B (014) Geometry of some taxicab
curves. Available at:

\textit{https://arxiv.org/ftp/arxiv/papers/1405/1405.7579.pdf}

Ma R, Sun E, Zou J (2023) A spectral method for assessing and combining
multiple data visualizations. \textit{Nature Communications}, 14(1):

780 doi :10.1038/s41467-023-36492-2

Ma R, Sun E, Zou J (2023) A spectral method for assessing and combining
multiple data visualizations. \textit{https ://arxiv.org/pdf/2210.13711.pdf}

Malinvaud E (1987) Data analysis in applied socio-economic statistics with
special consideration of correspondence analysis. In \textit{Marketing Science
Conference, Jouy en Josas}.

Meila M, Zhang H (2023) Maniold learning: what, how, and why. Available at:
\textit{https://arxiv.org/pdf/2311.03757.pdf}

Sah P, Fokou%
\'{}%
e E (2019) What do asian religions have in common ? an unsupervised text
analytics exploration. \textit{arXiv:1912.10847}

Vankadara LC, von Luxburg U (2018) Measures of distortion for machine
learning. In \textit{32nd Conference on Neural Information Processing Systems}
(NeurIPS 2018), Montr\'{e}al, Canada.

Saporta G, Tambrea N (1993) About the selection of the number of components in
correspondence analysis. Available at:

\textit{https://cedric.cnam.fr/\symbol{126}%
saporta/Saporta\_TambreaASMDA93.PDF}

Wilson J (2011) Taxi Cab Geometry with Technology: Some Exploration Materials.
Available at:

\textit{http://jwilson.coe.uga.edu/MATH7200/TaxiCab/TaxiCab.html}
\end{document}